\documentclass[a4paper]{jpconf}
\usepackage{graphicx}
\usepackage{amsmath,amssymb,amsfonts,amsthm}
\usepackage{mathrsfs,cleveref}

\usepackage{caption}
\usepackage{subcaption}
\usepackage{siunitx}
\usepackage{multicol}
\usepackage{booktabs}

\usepackage{graphicx}
\graphicspath{{Figures/}}

\usepackage[hyphens]{url}

\begin{document}
\title{Degradation-aware data-enabled predictive control of energy hubs}

\author{Varsha Behrunani$^{1,2}$, Marta Zagorowska$^{1}$, Mathias Hudoba de Badyn$^{1}$, Francesco Ricca$^{1}$, Philipp Heer$^{2}$ and John Lygeros$^{1}$}

\address{$^{1}$Automatic Control Laboratory, ETH Zurich, Switzerland}
\address{$^{2}$Urban Energy Systems Laboratory, Swiss Federal Laboratories for Materials Science and
Technology (Empa), Dübendorf, Switzerland}

\ead{bvarsha@ethz.ch}

\begin{abstract}
%In 2021, the energy use in buildings represented 30\% of global energy demand, and 45\% of the total energy demand in Switzerland. 
Mitigating the energy use in buildings, together with satisfaction of comfort requirements are the main objectives of efficient building control systems. Augmenting building energy systems with batteries can improve the energy use of a building, while posing the challenge of considering battery degradation during control operation. We demonstrate the performance of a data-enabled predictive control (DeePC) approach applied to a single multi-zone building and an energy hub comprising an electric heat pump and a battery. In a comparison with a standard rule-based controller, results demonstrate that the performance of DeePC is superior in terms of satisfaction of comfort constraints without increasing grid power consumption. Moreover, DeePC achieved two-fold decrease in battery degradation over one year, as compared to a rule-based controller. 
\end{abstract}

\section{Introduction}
In 2021, the energy use in buildings represented 45\% of the total energy demand in Switzerland~\cite{bfe2022energieverbrauch}. Mitigating the energy use in buildings, together with satisfaction of comfort requirements are the main objectives of efficient building control systems. It has been shown that improving energy storage in buildings by introducing batteries helps achieving these objectives~\cite{Battery_Rosewater2019}. However, the operation of a battery is affected by time, use or operating conditions and the battery may degrade. The goal of this work is to devise an efficient control system for buildings taking into account battery degradation.

% In 2021, the energy use in buildings represented 30\% of global energy demand~\cite{Name:iea_buidlings}, and 14\% of the total energy demand in Switzerland~\cite{bfe2022energieverbrauch}. Mitigating the energy use in buildings, together with satisfaction of comfort requirements are the main objectives of efficient building control systems. It has been shown that improving energy storage in buildings by introducing batteries helps achieving these objectives~\cite{Battery_Rosewater2019}. However, the operation of a battery is affected by time, use or operating conditions and the battery may degrade. The goal of this work is to devise an efficient control system for buildings taking into account battery degradation.   

It is important to ensure efficient operation of the controller that can take into account the degradation of the battery over time. 
In this context, Model Predictive Control (MPC) has been shown to reduce energy usage while maintaining comfort and operational constraints~\cite{drgovna2020all}. However, first principles models of buildings, and in particular of the effects of degradation, are costly to develop and difficult to maintain. Hence, to efficiently mitigate energy consumption over the whole lifetime of buildings, it is crucial to minimize degradation of the battery without an extensive modelling effort. In this work, we overcome the difficulties of first principles modelling by extending existing data-based approaches to capture the behaviour of a building affected by battery degradation.

Data-Enabled Predictive Control (DeePC) is used to investigate the performance of an energy hub comprising of a battery affected by degradation, and a heat pump. In contrast to classical MPC, DeePC computes an optimal control strategy for a linear time-invariant system using sufficiently rich input-output trajectories of the system. In this work, we extend the use of DeePC to long-term operation of building climate control considering nonlinear battery degradation. 

The paper is structured as follows. In Section \ref{sec:deepc}, we provide the theoretical background on DeePC, which outlines the basis for the problem formulation in Section \ref{sec:prob_statement}. Models for the building, battery, and heat pump dynamics are summarized in Section \ref{sec:Model}, and results from the simulations are discussed in Section \ref{sec:results}.

\section{Preliminaries on DeePC}
\label{sec:deepc}
Consider a discrete LTI system at time $k \in \mathbb {N}_0$:% with of order $n$, with $m$ inputs and $p$ outputs, as
\begin{equation}
\label{eq:LTI}
\begin{aligned}
 x_{k+1}=&{}A x_k+B u_k \\
 y_k=&{}C x_k+D u_k
\end{aligned}
\end{equation}
where $x_k \in \mathbb{R}^n$ is the state of the system, $u_k \in \mathbb{R}^m$ is the input vector, and $y_k \in \mathbb{R}^p$ is the output vector. The system matrices are $A \in \mathbb{R}^{n \times n}, B \in \mathbb{R}^{n \times m}, C \in \mathbb{R}^{p \times n}, D \in \mathbb{R}^{p \times m}$. Let $u_{\mathrm{d}} = (u_{\mathrm{d}}(1),...,u_{\mathrm{d}}(T_{\mathrm{d}}))\in\mathbb{R}^{T_{\mathrm{d}}m}$ and $y_{\mathrm{d}} = (y_{\mathrm{d}}(1),...,y_{\mathrm{d}}(T_{\mathrm{d}}))\in\mathbb{R}^{T_{\mathrm{d}}p}$ denote the input and output trajectory of length $T_{\mathrm{d}}$. Let $L$, $T_\mathrm{d} \in \mathbb{Z}_{\geq 0}$ and $T_\mathrm{d} \geq L$. The input trajectory $u_\mathrm{d} \in\mathbb{R}^{T_{\mathrm{d}}m}$ is called \emph{persistently exciting of order $L$} if the Hankel matrix 
 \begin{equation}
\mathscr{H}_L(u_\mathrm{d}):=\left[\begin{array}{cccc}
u_1 & u_2 & \cdots & u_{T_\mathrm{d}-L+1} \\
u_2 & u_3 & \cdots & u_{T_\mathrm{d}-L+2} \\
\vdots & \vdots & \ddots & \vdots \\
u_L & u_{L+1} & \cdots & u_{T_d}
\end{array}\right]
\end{equation}
is full rank. Following ~\cite{Data_Coulson2019}, we have that $T_{\mathrm{d}}\geq (m+1)(L + n) - 1$. DeePC uses Hankel matrices constructed from persistently-exciting inputs and corresponding outputs in lieu of a model of the form \eqref{eq:LTI} to find optimal trajectories of the system~\cite{Data_Coulson2019}. We consider Hankel matrices $\mathcal{H}_{T_{\mathrm{ini}}+T_{\mathrm{f}}}\left(u_{\mathrm{d}}\right)$ and $\mathcal{H}_{T_{\mathrm{ini}}+T_{\mathrm{f}}}\left(y_{\mathrm{d}}\right)$, and a partitioning thereof,
\begin{equation}
\left(\begin{array}{c}
U_{\mathrm{p}} \\
U_{\mathrm{f}}
\end{array}\right):=\mathcal{H}_{T_{\mathrm{ini}}+T_{\mathrm{f}}}\left(u_{\mathrm{d}}\right), \quad\left(\begin{array}{c}
Y_{\mathrm{p}} \\
Y_{\mathrm{f}}
\end{array}\right):=\mathcal{H}_{T_{\mathrm{ini}}+T_{\mathrm{f}}}\left(y_{\mathrm{d}}\right).
\end{equation}
The \emph{Fundamental Lemma} presented by \cite{Data_Coulson2019} states that if the system \eqref{eq:LTI} is controllable and $u_{\mathrm{d}}$ is persistently exciting of order $k+n$, then any sequence $\mathrm{col}(u_{\mathrm{ini}},y_{\mathrm{ini}},u,y$) $\in\mathbb{R}^{T_{\mathrm{ini}} + T_{\mathrm{f}}}$ is a trajectory of the system, if and only if there exists $g\in\mathbb{R}^{T_{\mathrm{d}} - T_{\mathrm{ini}} - T_{\mathrm{f}} - n + 1}$ such that
\begin{equation} \begin{pmatrix}U_{\mathrm{p}}^T & Y_{\mathrm{p}}^T & U_{\mathrm{f}}^T & Y_{\mathrm{f}}^T\end{pmatrix}^Tg = \begin{pmatrix}u_{\mathrm{ini}}^T & y_{\mathrm{ini}}^T & u^T & y^T\end{pmatrix}^T. \label{eq:non_param_mod}
\end{equation}

\section{DeePC for energy management }
\label{sec:prob_statement}
In this section the data-driven optimization problem for the optimal operation of the energy hub and the building is presented. The main objective is to minimize the energy consumption from the grid while satisfying operational constraints. We consider an energy hub comprising a heat pump and a battery that is used to supply the thermal demand of a five zone building. The measured output $y_{\mathrm{e}}\in\mathbb{R}^7$ includes the temperatures of all zones in the building, $y_{\mathrm{s}}\in\mathbb{R}^{5}$, the output power of the heat pump, $y_{\mathrm{h}}$, and the voltage of the battery, $y_{\mathrm{b}}$. The input vector $u_{\mathrm{e}}\in\mathbb{R}^{22}$ constitutes the control inputs $u_{\mathrm{s}}\in\mathbb{R}^9$ to the building, the power input to the heat pump $u_{\mathrm{h}}$, the battery current $u_{\mathrm{b}}$, and the disturbances, $v_{\mathrm{s}}\in\mathbb{R}^{11}$. The inputs $u^{i}_{\mathrm{s}}$, $i=1,\ldots,5$ describe the input to the radiators in each of the five zones, and $u^{i}_{\mathrm{s}}$, $i=6,\ldots,9$ describe the blinds openings available in four rooms. The disturbances $v_{\mathrm{s}}$ in are assumed to be known exactly from an accurate forecast. 

The DeepC controller computes the setpoints and we assume that low level controllers ensure that the setpoints are reached. The key advantage of DeePC is that we work directly with data, thus avoiding the need to model these low-level controllers. The resulting DeePC optimization for the optimal energy hub control over a prediction horizon $T_f$ is formulated as:
\begin{subequations}\label{eq:bldg_deePC}
\begin{align}
\label{cost_bldg}
\min_{u_{\mathrm{e}}, y_{\mathrm{e}},g} \quad\quad \ \ \sum\limits_{k=0}^{T_f-1}\left( \beta p^k +  \frac{1}{2\beta}c^k \right)^{2} &+  \lambda_{\rho}\left\| \rho\right\|_{2}^{2} +\lambda_g\|g\|_2^2 \\
\label{deepc_bldg} \textrm{s.t.} \quad \begin{pmatrix}U_{\mathrm{p}}^T & Y_{\mathrm{p}}^T & U_{\mathrm{f}}^T & Y_{\mathrm{f}}^T\end{pmatrix}^T g & = \begin{pmatrix}u_{\mathrm{e,ini}}^T & y_{\mathrm{e,ini}}^T & u_{\mathrm{e}}^T & y_{\mathrm{e}}^T\end{pmatrix}^T \\
\label{input1_bldg} u^{k}_{\mathrm{s,min}} &\leqslant u^{k,\mathrm{i}}_{\mathrm{s}} \leqslant u^{k}_{\mathrm{s,max}}, \ \ i=1\ldots,5  \\
 \label{output_bldg} y^{k}_{\mathrm{s,min}} - \rho &\leqslant y^{k}_{\mathrm{s}} \leqslant y^{k}_{\mathrm{s,max}} + \rho  \\ 
 \label{input_batt}u^{k}_{\mathrm{b, \min}} &\leqslant u^{k}_{\mathrm{b}} \leqslant u^{k}_{\mathrm{b, \max}}  \\
 \label{output_batt}y^{k}_{\mathrm{b, min}} &\leqslant y^{k}_{\mathrm{b}} \leqslant y^{k}_{\mathrm{b, max}} \\
\label{input_hp} 0 &\leqslant  y^{k}_{\mathrm{h}} \\
\label{output_hp} y^{k}_{\mathrm{h}} & = C_{\mathrm{h}} \cdot u^{k}_{\mathrm{h}}\\
\label{energy_hp} y^{k}_{\mathrm{h}} & = \sum_{i=1}^5 \frac{1}{\alpha_{i}} u^{k,\mathrm{i}}_{\mathrm{s}} \\
 \label{energy_bal} u^{k}_{\mathrm{h}} & = p^{k} + 0.066 \cdot u^{k}_{\mathrm{b}} 
\end{align}
\end{subequations} 
where $p^k$ is the energy imported from the electricity grid at time $k$ and $c^k$ is the price of energy imported from the grid. The cost function \eqref{cost_bldg} comprises of the linear cost of the electricity over the prediction horizon $T_{\mathrm{f}}$, in addition to the quadratic penalization on the power coming from the grid to improve numerical properties of the optimization problem with $\beta=0.01$. The cost also includes a regularization on the norm of $g$ with the parameter $\lambda_{g}=1000$ to avoid over-fitting and improve robustness~\cite{elokda2021data}. The slack $\rho$ on the comfort constraints is also penalised quadratically by the parameter $\lambda_{\rho}=10$.

The DeePC control strategy is incorporated in the constraint \eqref{deepc_bldg} in order to optimize the room temperatures in the buildings. The inequality constraints \eqref{input1_bldg} limit the radiator and blind inputs of the building, and \eqref{output_bldg} is the comfort constraint that ensures that the temperatures of the five zones remains within the time-dependent maximum and minimum temperatures, $y^{k}_{\mathrm{s,min}}$ and $y^{k}_{\mathrm{s,max}}$. The temperature is between $\SI{10}{\degree}$C and $\SI{40}{\degree}$C between 23:00 and 5:00 when the building is unoccupied, and between $\SI{21}{\degree}$C and $\SI{25}{\degree}$C during regular hours.  A slack variable $\rho$ on the comfort constraints ensures that the problem remains feasible for all disturbances $v_{\text{s}}^k$. The constraints \eqref{input_batt} and \eqref{output_batt} limit the current $u_b^k\in[-22,22]$ A and voltage $y_b^k\in[63,68]$ V of the battery so that the battery is charged or discharged at a maximum C-rate of $\frac{C}{4}$ based on the capacity of the battery, and that the battery voltage operates in the nominal region. The static model of the heat pump is incorporated using \eqref{input_hp} and \eqref{output_hp} with the coefficient of performance $C_h=3$, and \eqref{energy_hp} relates the output power from the heat pump to the heating power of the five radiators, where $\alpha_{i}$ are the coefficients corresponding to conversion factors with $\alpha_{i}=11.9$ for $i=1,2,3$, $\alpha_{4}=27.77$, $\alpha_{5}=7.58$. 

Finally, \eqref{energy_bal} is energy balance equation for the electricity in the hub, i.e. the power coming from the grid and the battery must be equal to the power going into the heat pump. Since both the voltage and current of the battery are decision variables, it would result in a bilinear equation that is difficult to solve. As a result, the battery voltage of 66V is used as an operating point in order to linearize this constraint.

% Table ?? specifies the inputs and output limits for the building and energy hub components used in this study.
% \textcolor{red}{table}

% \begin{table}[]
% \caption{}
% \label{tab:my-table}
% \begin{tabular}{lllllllllllllllll}
%       & $\beta$ & $\lambda_{\rho}$ & $\lambda_{g}$ & $u_{s,\min}^k$ & $u_{s,\max,i}^k,i=1,\ldots,5$ & $u_{s,\max,i}^k,i=6,\ldots,9$ & $y_{s,\min}^k$                     & $y_{s,\max}^k$                     & $u_{b,\min}^k$ & $u_{b,\max}^k$ & $y_{b,\min}^k$ & $y_{b,\max}^k$ & $C_h$ & $\alpha_{1,2,3}$ & $\alpha_{4}$ & $\alpha_{4}$ \\
% Value & 0.01    & 10               & 1000          & 0              & 70kW                          & 1                             & 10 (23:00-05:00), 21 (05:00-23:00) & 40 (23:00-05:00), 25 (05:00-23:00) & -50A           & 50A            & 63V            & 68V            & 3     & 11.90            & 27.77        & 7.58        
% \end{tabular}
% \end{table}

\section{Energy hub Modelling}
\label{sec:Model}
%The performance of DeePC is tested on a simulated building and energy hub model using \texttt{MATLAB}. 
%Detailed models of the building, the heat pump and the battery are created in \texttt{simulink}. 
The performance of DeePC is tested on a simulated building and energy hub model, created using the energy hub component modelling (EHCM) toolbox \cite{darivianakis2015ehcm} in \texttt{MATLAB}. %The models used are detailed below.
\subsection{Building}

In this work, we use an office building with five rooms (zones). The building is modelled using the Building Resistance Capacitance Modeling (BRCM) Toolbox which describes the building’s thermal dynamics as a continuous time system, bilinear in the inputs:% with a resistance-capacitance (RC) network. The resultant LTI system given by: 
\begin{equation}
\begin{aligned}
    \dot{x}_{\text{s}}(t)=&{}A_\text{c}x_{\text{s}}(t)+B_\text{u}u_{\text{s}}(t)+B_\text{v}v_{\text{s}}(t)+\sum\limits_{i=1}^9B_{\text{vu,i}}v_{\text{s}}(t)u_{\text{s}}(t)\\
    y_{\text{s}}(t)=&{}C_\text{c}x_{\text{s}}(t)
    \end{aligned}
\end{equation}
where the states of the systems, $x_{\mathrm{s}}\in\mathbb{R}^{113}$, include the temperatures of each room and the temperatures of the layers of the building elements i.e. floors, roof, and inner/outer walls that connect the zones of the building. The input vector $u_{\mathrm{s}}\in\mathbb{R}^{9}$ constitutes the control inputs, including the heating power of the radiators installed in the five zones ($\SI{}{\watt \meter}^2$), and inputs for the position of the four blinds on each facade of the building. The disturbance vector $v^{k}_{\mathrm{s}}\in\mathbb{R}^{11}$ comprises of the internal gains in the five rooms ($\SI{}{\watt \meter}^2$), ambient temperature ($\SI{}{\degree}$C), ground temperature ($\SI{}{\degree}$C), and the global solar radiation on the four sides of the building ($\SI{}{\watt \meter}^2$). The model thus considers external heat fluxes going into or coming out of the building including internal gains due to occupancy, lights and equipment, heating power from the radiators, and disturbances from ambient and ground temperature and heat gains by global solar radiation. A detailed description of the model and the matrices $A_c$, $B_u$, $B_v$, $B_{vu}$, $C_c$ can be found in \cite{darivianakis2015ehcm}.
% \begin{equation}
% \begin{aligned}
% & x^{k+1}_{\mathrm{s}}=A \cdot x^{k}_{\mathrm{s}}+B_{\text{u}}\cdot u^{k}_{\mathrm{s}}+B_{\text{v}} \cdot v^{k}_{\mathrm{s}} \\
% & y^{k}_{\mathrm{s}}=C x^{k}_{\mathrm{s}}
% \end{aligned}
% \end{equation}

\subsection{Battery}
A lithium-ion battery is considered for the energy hub modelled using a nonlinear Shepherd’s model, which describes how the terminal battery voltage changes with the input current~\cite{Battery_Mathworks2022}. The output of the model, $y^{k}_{\mathrm{b}}$, is the battery terminal voltage computed as $y^{k}_{\mathrm{b}}= V_{\mathrm{OC}}-R_0 \cdot u^{k}_{\mathrm{b}}$, where $u^{k}_{\mathrm{b}}$ is the battery current [A], $V_{\mathrm{OC}}$ is the open circuit voltage [V], and $R_0$ is the internal resistance of the battery [$\Omega$]. The internal resistance of the battery is affected by degradation. Battery degradation is modelled as ageing defined as the number of \emph{full cycles}, i.e. the number of times the State-of-Charge (SoC) goes from zero to 100\% and back to zero. The battery with parameters 12.8 V 40 Ah is implemented using the battery block in \texttt{Simulink} that includes the effects of cycling.%accounts for the cycling effects.% due to cycling.

\subsection{Heat pump}
Heat pump uses electricity to generated heat to satisfy the building heating demand. The EHCM toolbox uses a static model of the heat pump where the input and output at each time step related through the Coefficient of Performance (COP) as given in \eqref{output_hp}.
%\begin{equation}
%y^{k}_{\mathrm{h}} = \mathrm{COP} \cdot u^{k}_{\mathrm{h}}
%\\
%\label{eqn:hp_mod}
%\end{equation}

\section{Results}
\label{sec:results}
The controller from Section \ref{sec:prob_statement} has been implemented in the case study from Section \ref{sec:Model}. All simulations were performed in \texttt{MATLAB}/\texttt{Simulink} 2022a with \texttt{YALMIP} \cite{YALMIP_Lofberg2004} and \texttt{Gurobi} \cite{gurobi_2023}. The total simulation horizon was chosen as one year, $T_{\text{all}}=8760$ h. The controller was implemented in a receding horizon fashion, with prediction horizon of $T_f=24$ h. The parameter $T_{\text{ini}}=30$ h was chosen to minimise the prediction error between $Y_p$ and the true output over the prediction horizon $T_f$, at a fixed sampling time of $T_{\text{s}}=1$ h. The code is available in~\cite{gitlab_repo}

\subsection{Data collection}
The first step in solving the problem \eqref{eq:bldg_deePC} consists in collecting input and output data for the Hankel matrices in \eqref{deepc_bldg}. Measurements from the battery, the building and the heat pump were taken over $T_d=4416$ hours (184 days). The ageing effects on the battery were not taken into account in the data collection phase. For data collection, we used rule-based controllers (RBC) for the radiators and the blinds to ensure that the temperatures of the five zones stay within time-varying bounds:
\begin{equation}
   u_{\text{s}}^{k,i}=\delta^{k,i}_{\text{s}}+\begin{cases}
u^k_{s,\max} \text{ if }y_{\text{s}}^k\leq y^k_{s,\min}\\
u^k_{s,\min} \text{ if }y_{\text{s}}^k\geq y^k_{s,\max}
    \end{cases}
\end{equation}
where $\delta^{k,i}_{\text{s}}$ is an auxiliary input disturbance, chosen as a pseudo-random binary signal (PRBS) with amplitude of 5kW, necessary to ensure the condition on persistence of excitation. The battery controller is based on State-of-Charge (SoC) and the time of day. From midnight to 4 a.m. we charge the battery with a 15A current up until its SoC reaches 90\%. Then from 5 a.m. up until 23 p.m., the battery gets discharged. Then, when the SoC reaches 20\%, we wait for the next charge during the night. To ensure the persistence of excitation, the applied current is also perturbed with a PRBS signal with amplitude of 15A. The rule-based controllers were then used to find $u_{\text{ini}}$ and $y_{\text{ini}}$. 
%The RBC controller for the battery was designed by taking into account the higher electricity prices during the day. In this way, it is possible to satisfy the heating needs of the building during the day by drawing power from the battery instead of buying power from the grid.

% The amount of data collected for the Hankel matrices is defined by the value of $T_d$. The performance of deePC for nonlinear noisy systems improves if the Hankel matrix of the optimization problem \eqref{eq:bldg_deePC} has at least as many columns as rows, as it ensures that the subspace spanned by its columns contains the actual subspace of possible trajectories of the system \cite{Data_Elokda2021}. Therefore, $T_d$ should satisfy:
% \begin{equation}
%     T_d\geq\max\lbrace (m+1)(T_{\text{ini}}+T_{\text{f}}+n)-1,(m+p+1)(T_{\text{ini}}+T_{\text{f}})-1 \rbrace
% \end{equation}
% i.e. $T_d$ must be long enough so to satisfy the necessary condition for the persistency of excitation and to guarantee that the Hankel matrix of the deePC algorithm is square. 

% \textcolor{red}{line about DEEPC implementation - sampling time and explanation of MPC, amount of data collected prior to deepc }

\subsection{Prediction}
% Figure \ref{fig:AllComparison} and \ref{fig:BatteryCurrent} show the temperatures in the five rooms and the battery current input over the first two weeks of operation. During the data collection phase for $y_{ini}$ and $u_{ini}$, which lasted 30 hours, the temperatures on the five rooms oscillated between 20 and 28 $^{\circ}$ C. After the initial period, DeePC from \eqref{eq:bldg_deePC} kept the temperatures within the desired range. The battery current applied by DeePC (Fig. \ref{fig:BatteryCurrent}) was also less extreme than in the RBC case, which led to a less extreme behaviour of the energy hub.

\begin{figure}[!tbp]
\vspace{-0.3cm}
     \centering
         \includegraphics[width=0.99\textwidth]{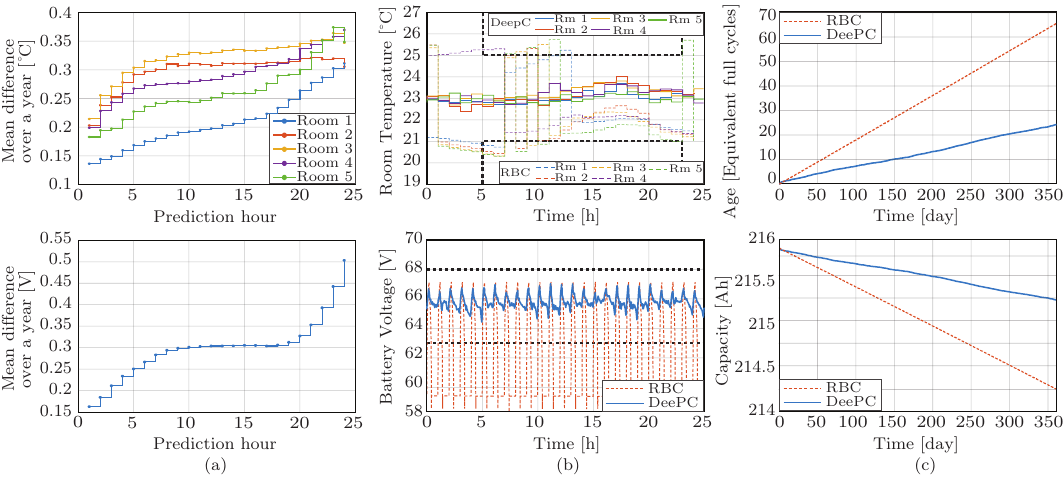}
         \vspace{-0.3cm}
        \caption{(a) Average prediction error for the five room temperatures and the battery voltage using DeePC over the entire year. (b) Comparison of the room temperature and the voltage output using DeepC and RBC over a selected day in January (24 h). (c) Evolution of age and capacity of the battery using DeePC and RBC over the entire year.}
        \label{fig:deepCall}
\vspace{-0.3cm}
\end{figure}

The choice of parameters was validated by evaluating the absolute error between the predicted outputs $y_{\text{s}}$ obtained from solving \eqref{eq:bldg_deePC} and the response of the building from simulation $y_{\text{s, sim}}$ at a given prediction hour, $k=1,\ldots,T_\text{f}$. Let $j=1,\ldots,5$ correspond to the room number, then the average error for each room for each prediction hour is given by 
$\epsilon^k_j=\sum_{i=1}^{T_{\text{all}}}|y^k_{\text{s},i,j}-y^k_{\text{s, sim},i,j}|/T_{\text{all}}$, where $T_{\text{all}}=8760$ h is the complete simulation time of one year. Figure \ref{fig:deepCall}(a) shows the error for the five rooms as a function of the prediction time. For all the rooms the average error is below 0.5 $^\circ$C which is considered acceptable \cite{Impact_Picard2017}. Moreover, the prediction error in the battery voltage remains below 0.5 V on average (Figure \ref{fig:deepCall})(a) which is below 1 \%. Figure \ref{fig:deepCall}(b) shows the output room temperatures and battery voltage obtained using DeepC and RBC over a selected day in January (24 h) and it shows how DeepC results in better temperature and voltage regulation.   

%\begin{figure}[!tbp]
%     \centering
%     % \hfill
%     \begin{subfigure}[b]{0.32\textwidth}
%         \centering
%         \includegraphics[width=\textwidth]{AverageErrorYear_big}
%        \caption{Average prediction error from DeePC for the five rooms, over the entire year}
%        \label{fig:PredictionHorizon}
%     \end{subfigure}
%               \begin{subfigure}[b]{0.32\textwidth}
%         \centering
%         \includegraphics[width=\textwidth]{VoltageError_big}
%        \caption{Average prediction error from DeePC for the battery voltage, over the entire year}
%        \label{fig:VoltageError}
%     \end{subfigure}
%               \begin{subfigure}[b]{0.32\textwidth}
%         \centering
%         \includegraphics[width=\textwidth]{A1_big}
%        \caption{Age and capacity of the battery if DeePC and RBC was used}
%        \label{fig:AgeComparison}
%     \end{subfigure}
%          % \hfill
%        \caption{DeePC performance for building application}
%        \label{fig:DeePCPrediction}
%\end{figure}

\subsection{Long-term operation}

Figure \ref{fig:deepCall}(c) shows the performance of the battery over the entire year. The oscillatory nature of the battery current in RBC led to intensification of battery ageing process quantified as the number of cycles (top) and loss of capacity (bottom). Even though the objective function from \eqref{cost_bldg} was focused only on the cost of operation of the energy hub, by optimising the input current to the battery and reducing its oscillatory nature, DeePC led to two times smaller age of the battery as the number of cycles, compared to RBC. The capacity loss was also reduced (0.3\% in DeePC compared to 0.8\% in RBC). Furthermore, DeePC enabled enforcing constraints on the battery voltage. Conversely, RBC adjusts only the battery current, which led to oscillatory behaviour contributing to the intensified ageing of the battery.

Quantitative results for constraint violation are collected in Table \ref{tbl:performanceComparison}. RBC violates the comfort constraints on average up to 5.5\% of the year whereas DeePC violates the constraints up to 2.8\% of the time and has smaller violations. At the same time, the cost is comparable in both RBC and DeePC, with DeePC cost being 0.9\% lower. 

\begin{table}[!tbp]
\caption{Performance comparison in terms of lower (LBV) and upper bound (UBV) constraint violations, and the overall cost.}
\vspace{-0.2cm}
\label{tbl:performanceComparison}
\centering
\scriptsize
\begin{tabular}{@{}p{1.1cm}p{2.6cm}p{2.6cm}p{2.5cm}p{2.5cm}p{2.1cm}@{}}
\toprule 
      & LBV/room/hour [$^{\circ}$C] & UBV/room/hour [$^{\circ}$C] & $\#$LBV/room [\% of time] & $\#$UBV/room [\% of time] & Cost {[}CHF{]} \\ \midrule
DeePC & 0.4                            & 0                               & 2.8                          & 0.2                     & 5909.8            \\
RBC   & 0.8                           & 0.2                            & 5.5                          & 3.5                           & 5961.7            \\ \bottomrule
\end{tabular}
\vspace{-0.5cm}
\end{table}

\section{Conclusions}
In this work, we have investigated the performance of data-enabled predictive control for building energy management through a simulation that incorporates degradation processes that affect battery behaviour. A simulation setup with a single building and an energy hub comprising an electric heat pump and a battery was considered. A comparison between DeePC and the RBC showed that the battery ageing was reduced by over a factor of two under DeePC operation, as well as a reduction of constraint violations. The impact of the simplified model of battery degradation (no calendar ageing, no self-discharge, no influence of varying external conditions) requires further investigation, ideally in an experimental setup. Future studies also aim to investigate the performance of DeePC as compared to more realistic rule-based controllers and on larger energy hubs with uncertain PV generation and the influence of the battery, as well as to extend the proposed approach to multiple buildings in a district. 

\section*{Acknowledgments}
Research supported by NCCR Automation, a National Centre of Competence in Research, funded by the Swiss National Science Foundation (grant no. 180545), and by the European Research Council (ERC) under the H2020 Advanced Grant no. 787845 (OCAL).

\vspace{-0.2cm}
\section*{References}
\bibliography{iopart-num}

\providecommand{\newblock}{}
\begin{thebibliography}{10}
\expandafter\ifx\csname url\endcsname\relax
  \def\url#1{{\tt #1}}\fi
\expandafter\ifx\csname urlprefix\endcsname\relax\def\urlprefix{URL }\fi
\providecommand{\eprint}[2][]{\url{#2}}
% Bibliography created with iopart-num v2.1
% /biblio/bibtex/contrib/iopart-num

\bibitem{bfe2022energieverbrauch}
{Bundesamt für Energie BFE} 2022 {Energieverbrauch nach Verwendungszweck}
  available: 4.04.2023
  \urlprefix\url{bfe.admin.ch/bfe/de/home/versorgung/statistik-und-geodaten/energiestatistiken.html}

\bibitem{Battery_Rosewater2019}
Rosewater D~M, Copp D~A, Nguyen T~A, Byrne R~H and Santoso S 2019 {\em IEEE
  Access\/} {\bf 7} 178357--178391

\bibitem{drgovna2020all}
Drgo{\v{n}}a J, Arroyo J, Figueroa I~C, Blum D, Arendt K, Kim D, Oll{\'e} E~P,
  Oravec J, Wetter M, Vrabie D~L {\em et~al.\/} 2020 {\em Annual Reviews in
  Control\/} {\bf 50} 190--232

\bibitem{Data_Coulson2019}
Coulson J, Lygeros J and Dörfler F 2019 Data-enabled predictive control: In
  the shallows of the {DeePC} {\em 2019 18th European Control Conference
  (ECC)\/} pp 307--312

\bibitem{elokda2021data}
Elokda E, Coulson J, Beuchat P~N, Lygeros J and D{\"o}rfler F 2021 {\em
  International Journal of Robust and Nonlinear Control\/} {\bf 31} 8916--8936

\bibitem{darivianakis2015ehcm}
Darivianakis G, Georghiou A, Smith R~S and Lygeros J 2015 {\em ETH Z{\"u}rich
  Automatic Control Laboratory\/}

\bibitem{Battery_Mathworks2022}
{The MathWorks Inc} 2022 Battery - generic battery model ({Simscape}
  electrical) available: 25.04.2023
  \urlprefix\url{ch.mathworks.com/help/sps/powersys/ref/battery.html}

\bibitem{YALMIP_Lofberg2004}
L{\"{o}}fberg J 2004 {YALMIP}: A toolbox for modeling and optimization in
  {MATLAB} {\em In Proceedings of the CACSD Conference\/} (Taipei, Taiwan)

\bibitem{gurobi_2023}
{Gurobi Optimization, LLC} 2023 {Gurobi Optimizer Reference Manual}
  \urlprefix\url{https://www.gurobi.com}

\bibitem{gitlab_repo}
Behrunani V, Zagorowska M, Hudoba~de Badyn M, Ricca F, Heer P and Lygeros J
  2023 {\em \url{http://dx.doi.org/10.3929/ethz-b-000615750}\/}

\bibitem{Impact_Picard2017}
Picard D, Drgoňa J, Kvasnica M and Helsen L 2017 {\em Energy and Buildings\/}
  {\bf 152} 739 -- 751 ISSN 0378-7788

\end{thebibliography}

\end{document}